# Structural investigations in BaFe$_{2-x}$Ru$_x$As$_2$ as a function of Ru and temperature

Shilpam Sharma[1], A. Bharathi[1], K. Vinod[1], C. S. Sundar[1], V. Srihari[3], Smritijit Sen[2], Haranath Ghosh[2], A.K. Sinha[2] and S. K. Deb[2]

[1]Materials Science Group, Indira Gandhi Centre for Atomic Research, Kalpakam, India

[2]Raja Rammana Centre Advanced Technology, Indore, India

[3]Photon Factory, KEK, Tsukuba, Japan

**Abstract:**

We present the results of Synchrotron XRD measurements on powdered single crystal samples of BaFe$_{2-x}$Ru$_x$As$_2$, as a function of Ru content, and as a function of temperature, across the spin density wave transition in BaFe$_{1.9}$Ru$_{0.1}$As$_2$. The Rietveld refinements reveal that with Ru substitution, while the *a*-axis increases, the *c*-axis decreases. In addition the variation of positional co-ordinates of As ($z_{As}$), the Fe-As bond length and the As-Fe-As bond angles have also been determined. In the sample with *x=0.1*, temperature dependent XRD measurements, indicate that the orthorhombicity shows the characteristic increase with decrease in temperature, below the magnetic transition. It is seen that the c-axis, the As-Fe-As bond angles, Fe-As bond length and positional co-ordinate of the As show definite anomalies close to the structural transition. The observed anomalies in structural parameters are analysed in conjunction with geometric optimization of the structure using *ab*-initio electronic structure calculations.

# 1. Introduction

The common feature in the Fe-pnictide superconductors is the presence of FeAs plane, wherein the Fe-atoms form a regular square lattice, and the arsenic ions are located above/below the plane of Fe square. This arrangement of arsenic ions has several important consequences on the electronic and magnetic and superconducting properties of these systems (Yildirim, 2009, Johnston, 2010). Since As is not directly between two Fe ions, the Fe–Fe distance is not large and direct Fe–Fe overlap plays an important role in the band formation near the Fermi level. Then, the delicate interplay between Fe–Fe, Fe–As, and even As–As interactions result in interesting electronic and magnetic properties that are sensitive to the As z-position (Yildirim, 2009). Extensive studies on the Fe-arsenide superconductors have established that the parent $BaFe_2As_2$ undergoes a structural transition from the high temperature tetragonal to a low-temperature orthorhombic phase, associated with a Spin Density Wave (SDW) state transition, and superconductivity in this system is induced by suppressing the SDW state by introducing carriers through chemical doping at the Fe /As site or by the application of pressure (Canfield & Bud'ko, 2010, Johnston, 2010).

In this paper, we report on the results of in-depth structural studies on Ru- substituted $BaFe_2As_2$, wherein superconductivity has been obtained through iso-electronic substitution (Sharma *et al.*, 2010). The origin of superconductivity in this iso-electronically substituted system has attracted considerable attention, and a variety of studies have been carried out. These include, investigations on the changes in lattice parameters, in particular the development of orthorhombicity that is correlated with the magnetic transition and superconducting transition (Kim *et al.*, 2011); investigations on the dilution of magnetism at the Fe site, induced by Ru substitution, using X-ray Resonant Magnetic Scattering, (Kim *et al.*, 2013) and Mossbauer Spectroscopy (Reddy *et al.*); investigations on the changes in electronic structure with Ru doping and temperature using Angle Resolved Photo Emission Spectroscopy (Brouet *et al.*, 2010, Xu *et al.*, 2012), and studies on the exploration of the similarities between superconductivity induced by iso-electronic Ru doping and pressure induced superconductivity in parent $BaFe_2As_2$ (Devidas *et al.*, 2014). Dhaka et al have (Dhaka *et al.*, 2013) pointed out that there are significant changes, in the Fermi surface characteristics for both the pristine and Ru doped samples, as function of temperature. This they argue could be the reason for the occurrence of nesting at low temperatures leading to the SDW to paramagnetic phase transition. Further, they point out that this is possible only if the structural parameters change with temperature and show using the thermal expansion data and band structure calculations that the positional co-ordinate of the As atom changes with increase in temperature (Dhaka *et al.*, 2013). They however, did not have temperature dependent crystallographic data detailing the changes in ionic positions with temperature in arriving at their conclusions. Since the structural, electronic and magnetic properties of the system are intimately related, it is easy to appreciate the necessity of having detailed information on the variation of structural parameters both as a function of Ru substitution and as a function of temperature.

Here we determine the structural parameters of interest from Synchrotron XRD measurements in powdered single crystals of the $BaFe_{2-x}Ru_xAs_2$ system as a function of Ru concentration and as function of temperature across the magnetic transition for the $BaFe_{1.9}Ru_{0.1}As_2$ sample. In addition to obtaining information on the variation of lattice parameters, Rietveld analysis of the XRD data have been used to obtain detailed information on the variation of positional co-ordinate of the As atom ($z_{As}$), the Fe-Fe, and Fe-As bond distances and the Fe-As-Fe tetrahedral angle, which show anomalous variations in the vicinity of magnetic transition. It is shown that these anomalous variations of structural parameters, are also reflected in the results of geometric simulations based on ab-initio electronic structure calculations.

## 2. Experimental Details

Single crystals of $BaFe_{2-x}Ru_xAs_2$ were prepared without flux, by taking stoichiometric quantities of Ba, FeAs and RuAs in alumina crucibles and vacuum sealing them in quartz tubes, as described elsewhere (Shilpam *et al.*, 2013). The samples were heated at the rate of 50°C/hour until a temperature of 1190 °C at which they were soaked for 24 hours. The samples were slowly cooled up to 800°C at 1.5°C/hour and then furnace cooled to room temperature. Tiny single crystals typically 1mm x 1mm x 0.1mm were seen to form, which could be easily retrieved. The XRD patterns obtained from flat crystal surfaces show only *(00l)* reflections, suggesting that the crystal surfaces are perpendicular to the c-axis. XRD measurements at room temperature, were performed on the powdered single crystals in the BL-12 beam line, using 16 keV x-rays (Sinha *et al.*, 2011). The powdered samples were placed in a circular depression made on Kapton tape. The data were collected on a MAR3450 image plate detector in the transmission geometry. Fit2D program (Hammersley *et al.*, 1996) was employed to convert the 2D image from the image plate detector to 1D Intensity versus 2θ plot.

Temperature dependence of the structural parameters across the magnetic transition have been carried out for $BaFe_{1.9}Ru_{0.10}As_2$ using powder XRD using a point detector at 16 keV in the BL-18B beam line of Photon Factory in KEK, Tsukuba. The SDW transition temperature was established through resistivity measurements as function of temperature (4K to 300K) in the van der Pauw geometry, and was found to be consistent with the known phase diagram (Vinod *et al.*, 2012, Rullier-Albenque *et al.*, 2010, Thaler *et al.*, 2010). For the XRD measurements, finely powdered single crystals were spread in the copper cold finger of the cryostat along with grease. The measurements were done in the θ - 2θ geometry with a Cyberstar point detector, with a step size of 0.003 deg. The temperature variation from 300K to 20K was achieved with a closed cycle refrigerator. The temperature was stabilized to ±0.1 K at a given temperature using a Cryocon 32B temperature controller. The 2θ scans at each temperature, was carried out in the 13 - 40 degree 2θ range. The dwell time at each 2θ angle was 3 seconds. The highest peak count for each temperature was ~10000. The Rietveld refined fits of the data at each temperature were carried out with the GSAS program (Toby, 2001, Larson & Von Dreele, 1994). From the fits the values Fe-As bond lengths, the As position $z_{As}$ and As-Fe-As bond angles were obtained.

## 3. Results and discussions:

### 3.1. Variation in structure due to Ru substitution

The room temperature diffraction patterns obtained in powdered single crystals are shown in figure 1, for representative Ru fractions, along with the Rietveld fits. The refinements were carried out using the GSAS program (Larson & Von Dreele, 1994, Toby, 2001) with Pseudo Voigt function as peak profiles. It is evident from the difference plot shown in the figures, that the fits are very good. The lattice parameter variations as a function of Ru content 'x', obtained from the Rietveld fits of the room temperature data, are presented in figure 2. It is seen that with the increase of Ru content, the *a*-lattice parameter increases and the c-lattice parameter decreases, consistent with earlier results (Thaler *et al.*, 2010, Rullier-Albenque *et al.*, 2010, Eom *et al.*, 2012). In figure 3, we show the variation of the structural parameters with Ru content, viz., $z_{As}$, the position of the As along the c-direction, Fe-As bond lengths, and Fe-Fe distances and the two tetrahedral angles (see inset), as extracted from the results of the Rietveld refinements. It is seen that with the increase of Ru content, there is a decrease in $z_{As}$ and an increase in the Fe-Fe bond length, which are consistent with the observed decrease in the c- and an increase in a-lattice parameter respectively (cf. figure 2). The systematic decrease of $z_{As}$ with Ru substitution, is associated with an increase in the Fe-As bond lengths and the corresponding variation of the tetrahedral bond angles is indicative of a distortion of the FeAs$_4$ tetrahedra on account of Ru substitution. Similar observations on the decrease of $z_{As}$ and an increase in the tetrahedral bond angles with Ru substitution has been reported through studies on single crystalline samples by Rullier–Albenque et. al (Rullier-Albenque *et al.*, 2010).

While apriori, an increase in the *a*-lattice parameter (cf. figure 2) may reflect the larger size of Ru ion, it is difficult to rationalize the associated decrease in c lattice parameter. As first indicated by Yildrim (Yildirim, 2009), in the Fe-pnictides, the strength of interaction between arsenic ions in iron pnictides is controlled by the Fe-spin state. In particular, it has been shown in CaFe$_2$As$_2$ system that reducing the Fe-magnetic moment weakens the Fe-As bonding, and in turn, increases As-As interactions, causing giant reduction in the c axis. In the case of Ru doped BaFe$_2$As$_2$, spin polarized band structure calculations (Sharma *et al.*, 2010) indicate that the *d*-electron bands get delocalized with the addition of Ru 4d electrons, and this has the effect of decrease in the moment at the Fe site. The results from band structure calculations are shown in the inset of figure 3, and evidence for the decrease in moment at Fe site is also seen in the experiments such as XRMS and Mossbauer spectroscopy (Kim *et al.*, 2013, Reddy *et al.*). The results in figure 3 provide one more supportive evidence, through results in Ru doped $BaFe_2As_2$, on the influence of magnetic moment at the Fe site on the structural parameters. Given this, it is of interest to follow the changes in lattice parameter, across the magnetic transition, and in the following this has been done for $BaFe_{1.9}Ru_{0.1}As_2$, that has a SDW transition at 125 K.

### 3.2. Variation in structure across the SDW transition in $BaFe_{1.9}Ru_{0.1}As_2$

Figure 4 shows the Rietveld analysis of XRD for two representative temperatures, viz., 300 K and 50 K. Figure 5 summarizes the results of variation of resistivity and lattice parameters as a function of temperature across the SDW transition in $BaFe_{1.9}Ru_{0.1}As_2$. The sharp fall in resistivity at 122.8 K (cf. figure 5a) is associated with the spin density wave transition (Rotter *et al.*, 2008). It is seen that the derivative plot of resistivity shows only a single anomaly implying that the structural and magnetic transitions are concurrent in this system, unlike that in the Co doped $BaFe_2As_2$ samples (Canfield *et al.*, 2009). The variation of lattice parameters across the SDW transition, as obtained from Rietveld analysis of XRD data as a function of temperature (cf. figure 4) is shown in figure 5c, and the evolution of orthorhombic splitting, viz., (a-b)/(a+b) with the lowering of temperature is shown in figure 5b. The emergence of the orthorhombic phase, coincident with the SDW transition is consistent with earlier results (Kim *et al.*, 2011). Further, the degree of orthorhombic splitting as shown in figure 5b is identical to that obtained from an independent study on the temperature dependent splitting of the (110) peaks obtained from high resolution XRD measurements on single crystals, using the Petra-3 beam line (Bharathi *et al.*, 2012).

It is seen from figure 5c that in addition to the development of the orthorhombic splitting, there is an anomalous increase in the *c*-lattice parameter. Such anomalies in the *c*-lattice parameter and cell volume, have been reported across the tetragonal to orthorhombic phase transition, in $SrFe_2As_2$ and $EuFe_2As_2$ systems (Marcus *et al.*, 2008). The results of variation of various structural parameters, $z_{As}$, the Fe-Fe and Fe-As bond lengths and the As-Fe-As bond angles, as obtained from Rietveld refinements of the XRD data at different temperatures, are shown in figure 6. The following points are noteworthy: (1) The Fe-Fe bond length splits into short and long bonds in the orthorhombic phase (2) Anomalous variation in $Z_{As}$ and correspondingly in Fe-As bond lengths and the As-Fe-As bond angles are seen in the vicinity of the structural transition. While such anomalous variations in the structural parameters have not been reported so far, it must be mentioned that anomalous variations have been seen in the experiments on measurement of phonon frequencies in the vicinity of structural / magnetic phase transitions in BaFe$_2$As$_2$ (Rahlenbeck *et al.*, 2009, Teng *et al.*, 2013, Zhang & Zhang, 2012).

To check the veracity of the observations on the anomalous variations of structural parameters, obtained through Rietveld analysis, we have carried out first principle geometry optimization, using the material studio 6.1 CASTEP package (Clark Stewart *et al.*, 2005). We know that in the Ba-122 structure, only the position of As, $z_{As}$, can be changed for the geometry optimization, since both Fe and Ba atoms occupy symmetry positions in the structure I4/mmm (space group number No. 139). In these calculations, the variation of structural parameters of BaFe$_{2-x}$Ru$_x$As$_2$ as a function of temperature for x=0.1 and also as a function of Ru content at room temperature, are determined using the experimentally measured lattice parameters as inputs. The calculations are done within the generalized gradient approximation (GGA) using Perdew-Burke-Enzerhof (PBE) functional (Perdew *et al.*, 1996). Ru substitution in place of Fe is tackled by considering the mixed atoms based on the Mixture Atom Editor of CASTEP program in Material Studio CASTEP by forming Fe$_{2-x}$Ru$_x$ compound atom. The composition of Fe and Ru at the given site is modified so as to be consistent with Ru fraction substituted x=0.1, 0.2, 0.3 *etc*. Spin polarized geometry optimizations are carried out using anti-ferromagnetic spin-stripe configuration for the low temperature orthorhombic phase (Aktürk & Ciraci, 2009). The parameters of the calculations are the plane wave basis set with energy cut off 380 eV and SCF tolerance 10$^{-6}$ eV/atom. Brillouin zone is sampled in the k space within Monkhorst-Pack scheme and grid size for SCF calculation is (6x6x2).

The results of the above calculations, on the temperature dependence of structural parameters for BaFe$_{1.9}$Ru$_{.1}$As$_2$ are shown in figure 7. It is noted that the calculated Fe-Fe distances matches quantitatively with experimentally measured Fe-Fe distances shown in figure 6. Whereas, the calculated variation of $z_{As}$ is indicative of an anomalous temperature dependence behavior, it is not in quantitative agreement with the experimental results shown in figure 6c, and there is a large discrepancy between the calculated Fe-As and in particular the As-Fe-As bond angles and the results from experiments. As a next step, we have used the experimentally measured $z_{As}$ (cf. figure 6) along with the lattice parameters (cf. figure 5) into the calculation and extract the other structural parameters, viz., Fe-As bond length and As-Fe-As bond angles, and these results are shown in the left panel of figure 8. It is now seen that the variations of both the Fe-As bond length and the As-Fe-As bond angles with temperature track that measured by the experiment. This points to the importance of $z_{AS}$ in determining the electronic structure and consequently, the structural parameters of the Fe-pnictides. The results in the right panel of figure 8 show the results of the geometry optimization calculations on the variation of structural parameters as a function of Ru content. Comparing with the experimental results (cf. figure 3), we see that the important trends, viz., the increase of Fe-Fe distance with Ru doping and the distortion of the FeAs$_4$ tetrahedra as reflected in the divergence of the two tetrahedral angles are reproduced.

## 4. Summary and conclusions

In summary, synchrotron measurements have been carried out to investigate in detail the changes in structural parameters with Ru doping and as a function of temperature across the SDW transition. The variation in structural parameters, in particular the decrease in $z_{As}$ with Ru doping is correlated with the dilution of magnetic moment at the Fe site. Studies on the variation of structural parameters with temperature across the SDW transition, indicate in addition to the well-known development of orthorhombic splitting, an anomalous increase in c-lattice parameter, related to the anomalous variation in $z_{As}$ is observed. An optimization of structural parameters, have been carried out using electronic structure calculations (CASTEP-6.1) and these in turn point to the importance of magnitude of $z_{As}$. In particular, it is shown that with only $z_{As}$ taken from experimental results, the observed variation of structural parameters as Fe-Fe, Fe-As bond lengths, and tetrahedral bond angle with temperature and composition can be accounted for. We believe the structural data from these studies can be used as input for in-depth band structure calculations to compare with temperature dependent ARPES data (Dhaka *et al.*, 2013) and would lead further insight into the correlation between structure, electronic properties and magnetism in Fe-based superconductors.

## References

Aktürk, E. & Ciraci, S. (2009). *Physical Review B* **79**, 184523.
Bharathi, A., Vinod, K., Sharma, S. & Seeck, O. H. (2012). *Annual Report Photon Science - HASYLAB*.
Brouet, V., Rullier-Albenque, F., Marsi, M., Mansart, B., Aichhorn, M., Biermann, S., Faure, J., Perfetti, L., Taleb-Ibrahimi, A., Le Fèvre, P., Bertran, F., Forget, A. & Colson, D. (2010). *Physical Review Letters* **105**, 087001.
Canfield, P. C. & Bud'ko, S. L. (2010). *Annual Review of Condensed Matter Physics* **1**, 27-50.
Canfield, P. C., Bud'ko, S. L., Ni, N., Yan, J. Q. & Kracher, A. (2009). *Physical Review B* **80**, 060501.
Clark Stewart, J., Segall Matthew, D., Pickard Chris, J., Hasnip Phil, J., Probert Matt, I. J., Refson, K. & Payne Mike, C. (2005). Zeitschrift für Kristallographie, **220**, 567.
Devidas, T. R., Mani, A., Sharma, S., Vinod, K., Bharathi, A. & Sundar, C. S. (2014). *Solid State Communications* **185**, 62-66.
Dhaka, R. S., Hahn, S. E., Razzoli, E., Jiang, R., Shi, M., Harmon, B. N., Thaler, A., Bud'ko, S. L., Canfield, P. C. & Kaminski, A. (2013). *Physical Review Letters* **110**, 067002.
Eom, M. J., Na, S. W., Hoch, C., Kremer, R. K. & Kim, J. S. (2012). *Physical Review B* **85**, 024536.

Hammersley, A. P., Svensson, S. O., Hanfland, M., Fitch, A. N. & Hausermann, D. (1996). *High Pressure Research* **14**, 235-248.
Johnston, D. (2010). *Advances in Physics* **59**, 803-1061.
Kim, M. G., Pratt, D. K., Rustan, G. E., Tian, W., Zarestky, J. L., Thaler, A., Bud'ko, S. L., Canfield, P. C., McQueeney, R. J., Kreyssig, A. & Goldman, A. I. (2011). *Physical Review B* **83**, 054514.
Kim, M. G., Soh, J., Lang, J., Dean, M. P. M., Thaler, A., Bud'ko, S. L., Canfield, P. C., Bourret-Courchesne, E., Kreyssig, A., Goldman, A. I. & Birgeneau, R. J. (2013). *Physical Review B* **88**, 014424.
Larson, A. C. & Von Dreele, R. B. (1994). *Los Alamos National Laboratory Report LAUR* **86**, 748.
Marcus, T., Marianne, R., Veronika, W., Falko, M. S., Rainer, P. & Dirk, J. (2008). *Journal of Physics: Condensed Matter* **20**, 452201.
Perdew, J. P., Burke, K. & Ernzerhof, M. (1996). *Physical Review Letters* **77**, 3865-3868.
Rahlenbeck, M., Sun, G. L., Sun, D. L., Lin, C. T., Keimer, B. & Ulrich, C. (2009). *Physical Review B* **80**, 064509.
Reddy, R. V., Bharathi, A., Gupta, A., Sharma, K., Chandra, S., Sharma, S., Vinod, K. & Sundar, C. S. *Under Review*.
Rotter, M., Tegel, M., Johrendt, D., Schellenberg, I., Hermes, W. & Pöttgen, R. (2008). *Physical Review B* **78**, 020503.
Rullier-Albenque, F., Colson, D., Forget, A., Thuéry, P. & Poissonnet, S. (2010). *Physical Review B* **81**, 224503.
Sharma, S., Bharathi, A., Chandra, S., Reddy, V. R., Paulraj, S., Satya, A. T., Sastry, V. S., Gupta, A. & Sundar, C. S. (2010). *Physical Review B* **81**, 174512.
Shilpam, S., Vinod, K., Sundar, C. S. & Bharathi, A. (2013). *Superconductor Science and Technology* **26**, 015009.
Sinha, A. K., Sagdeo, A., Gupta, P., Kumar, A., Singh, M. N., Gupta, R. K., Kane, S. R. & Deb, S. K. (2011). *AIP Conference Proceedings* **1349**, 503-504.
Teng, J., Chen, C., Xiong, Y., Zhang, J., Jin, R. & Plummer, E. W. (2013). *Proceedings of the National Academy of Sciences* **110**, 898-903.
Thaler, A., Ni, N., Kracher, A., Yan, J. Q., Bud'Ko, S. L. & Canfield, P. C. (2010). *Physical Review B* **82**, 014534.
Toby, B. (2001). *Journal of Applied Crystallography* **34**, 210-213.
Vinod, K., Sharma, S., Satya, A. T., Sundar, C. S. & Bharathi, A. (2012). *AIP Conference Proceedings* **1447**, 889-890.
Xu, N., Qian, T., Richard, P., Shi, Y. B., Wang, X. P., Zhang, P., Huang, Y. B., Xu, Y. M., Miao, H., Xu, G., Xuan, G. F., Jiao, W. H., Xu, Z. A., Cao, G. H. & Ding, H. (2012). *Physical Review B* **86**, 064505.
Yildirim, T. (2009). *Physica C: Superconductivity* **469**, 425-441.
Zhang, Q. M. & Zhang, A. M. (2012). *Modern Physics Letters B* **26**, 1230020.

**Figure Captions**

**Figure 1:** The room temperature diffraction patterns obtained in powdered single crystals are shown for representative Ru fractions, for x=0.0 and x=0.2. The Rietveld refinements were carried out using the

GSAS program with Pseudo Voigt function for peak profiles, and the resultant fits are shown in figure as solid lines.

**Figure 2:** The lattice parameter variations as a function of Ru content, x, obtained from the Rietveld fits of the room temperature data. The c-lattice parameter decreases and while the a-lattice parameter increases with Ru substitution

**Figure 3**: The variation of structural parameters with Ru doping as obtained through Rietveld analysis, showing the decrease in $z_{As}$, the increase in the Fe-As and Fe-Fe bond lengths, and the increased divergence of the two tetrahedral bond angles, the latter shown in the FeAs plane of the structure, in the inset of panel (b). The decrease in the moment at the Fe site with Ru fraction, x, substituted, obtained from band structure calculations (Shilpam *et al.*, 2010) is shown in the inset of figure 3a.

**Figure 4** Rietveld refinement in the $BaFe_{1.9}Ru_{0.1}As_2$ sample on the data measured at room temperature and 50 K. The Rietveld refined fits shown as solid lines were carried out using the GSAS program with Pseudo Voigt function as peak profiles.

**Figure 5**: (a) Variation of resistance with temperature and temperature derivative of resistance showing the sharp SDW transition in $BaFe_{1.9}Ru_{0.1}As_2$. The orthorhombic splitting, (a-b)/(a+b), across the SDW transition is shown in figure (b). **Figure (c):** The temperature dependence of lattice parameters shows anomalous increase in *c*-lattice parameter below orthorhombic transition.

**Figure 6** Variation of structural parameters of $BaFe_{1.9}Ru_{0.1}As_2$ with temperature across the SDW transition. The development of two Fe-Fe bond lengths in the orthorhombic phase and anomalous temperature variations in $z_{As}$, As-Fe-As bond angles, Fe-As bond lengths are clearly seen**.**

**Figure 7:** Results of the first principles geometry optimization studies on the variation of structural parameters as a function of temperature in $BaFe_{1.9}Ru_{0.1}As_2$. It is seen that while the Fe-Fe bond length and $z_{As}$ are in conformity with the experimental results in figure 5, there is considerable discrepancy in Fe-As bond length and the As-Fe-As bond angles. (See text)

**Figure 8:** (Left Panel) Computed temperature dependence of structural parameters, viz., Fe-Fe, Fe-As bond lengths and As-Fe-As bond angles, with $z_{As}$ taken from experiments (see text) for $BaFe_{1.9}Ru_{0.1}As_2$. (Right panel) Structural parameters with the variation of Ru content, as obtained from ab-initio calculations

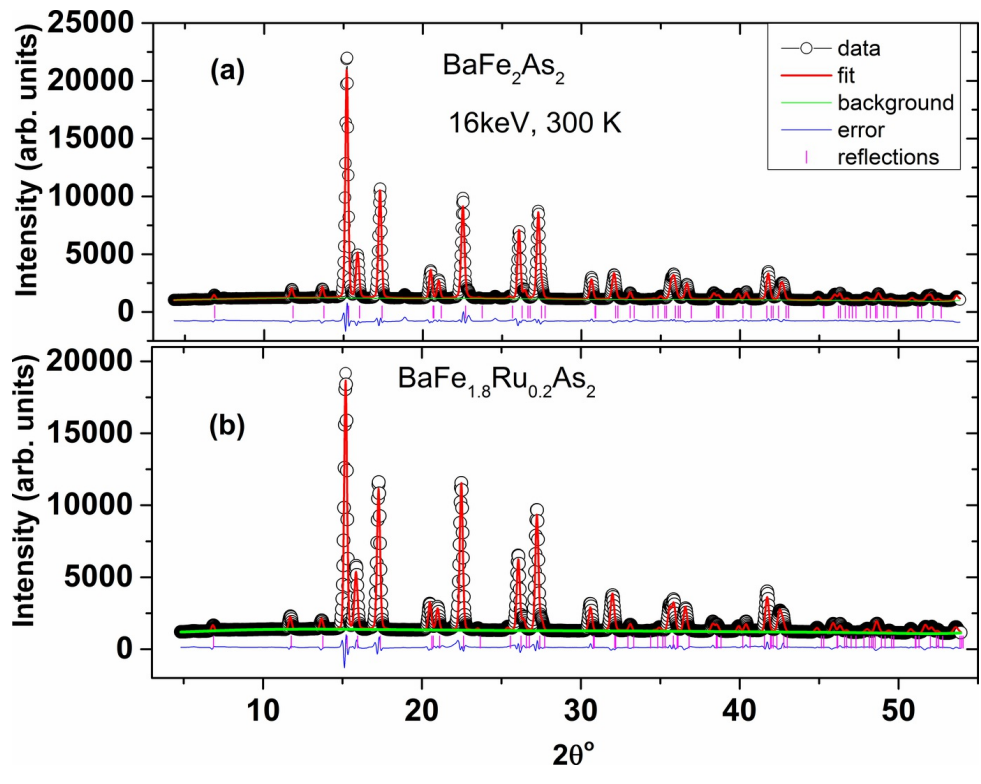

**Figure 1**

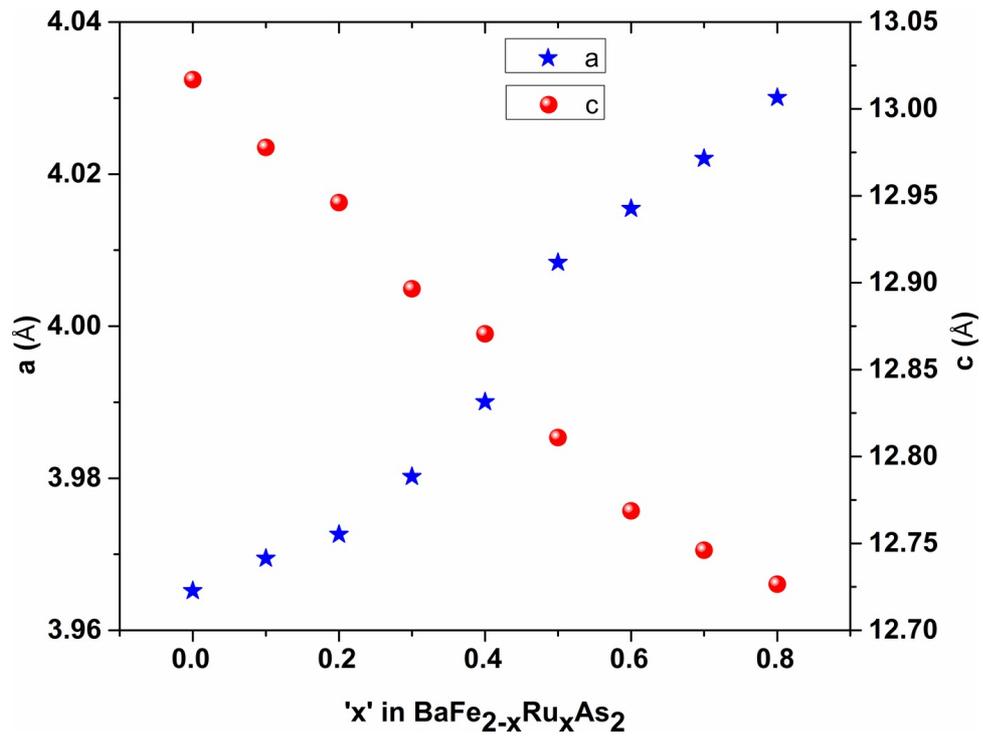

**Figure 2**

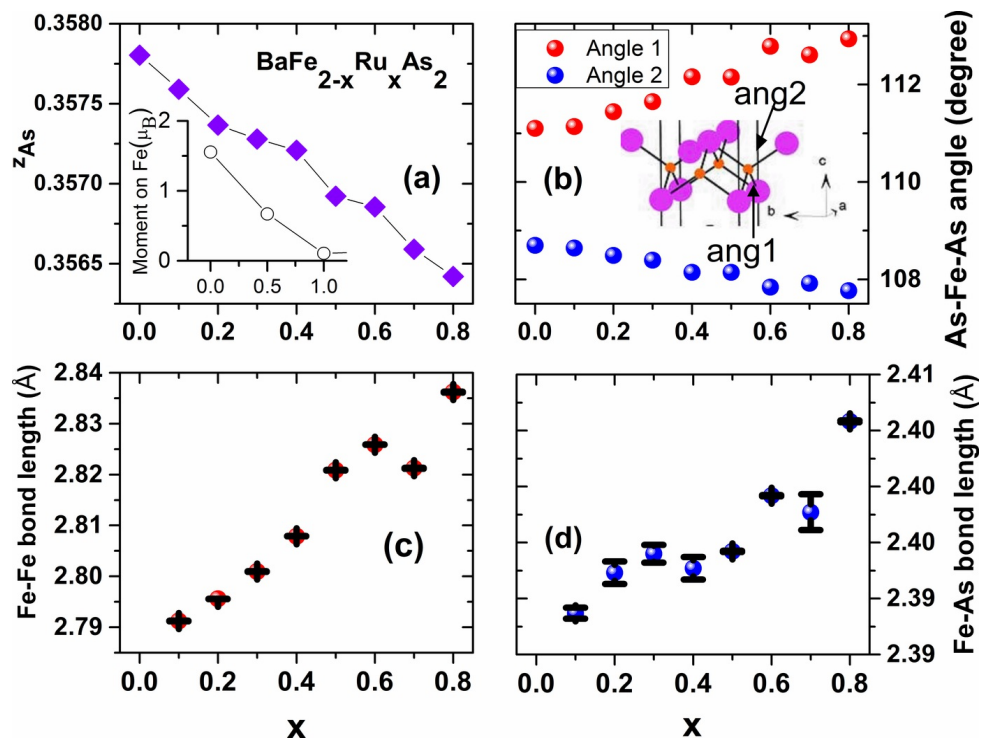

Figure 3

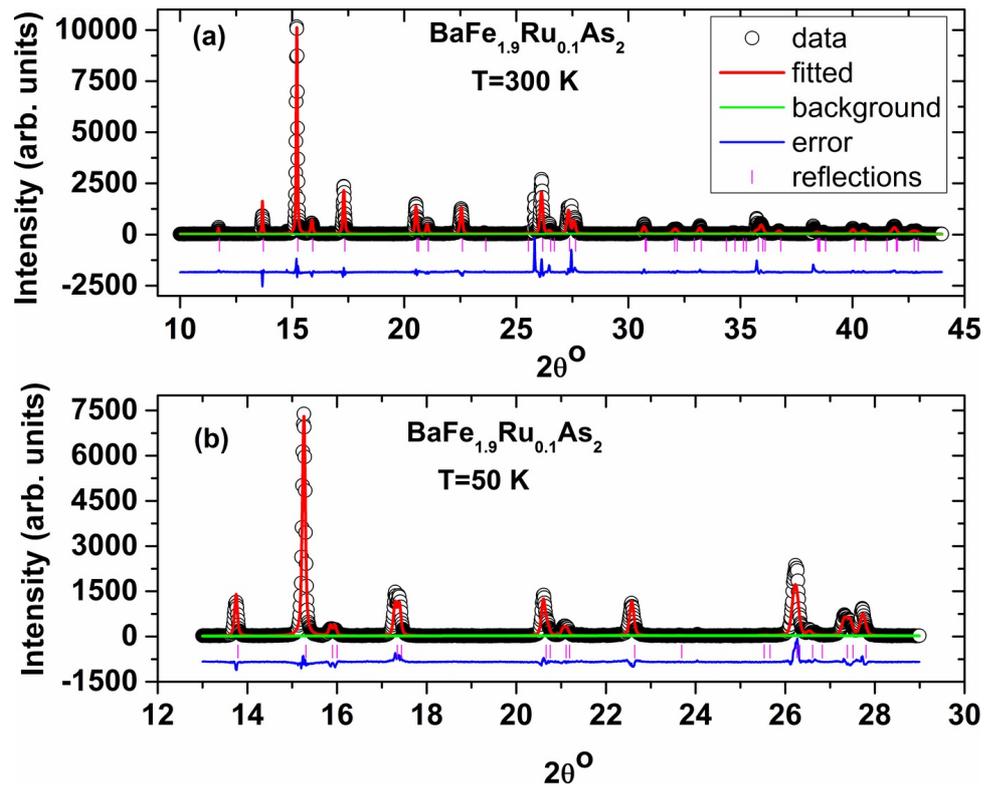

**Figure 4**

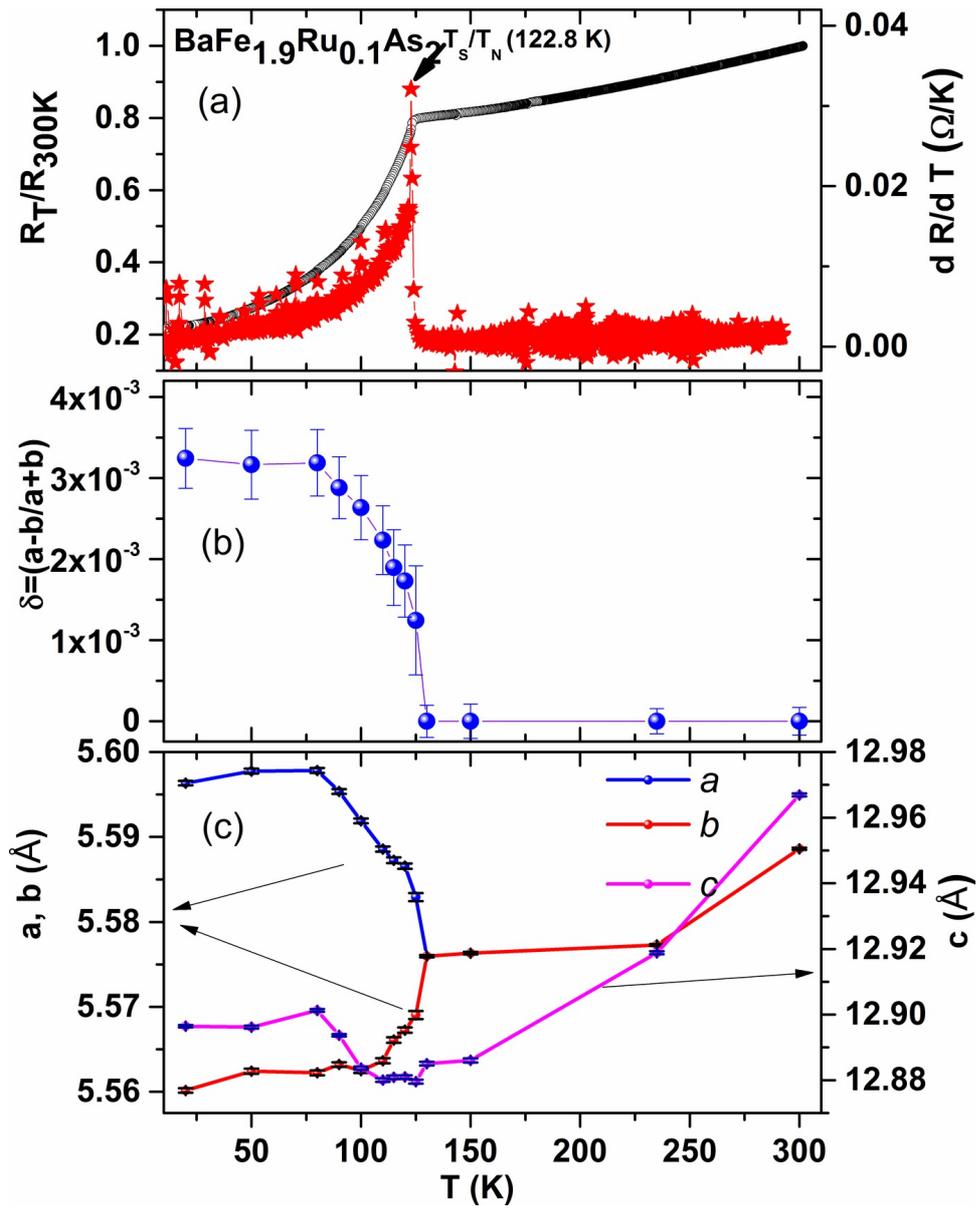

Figure 5

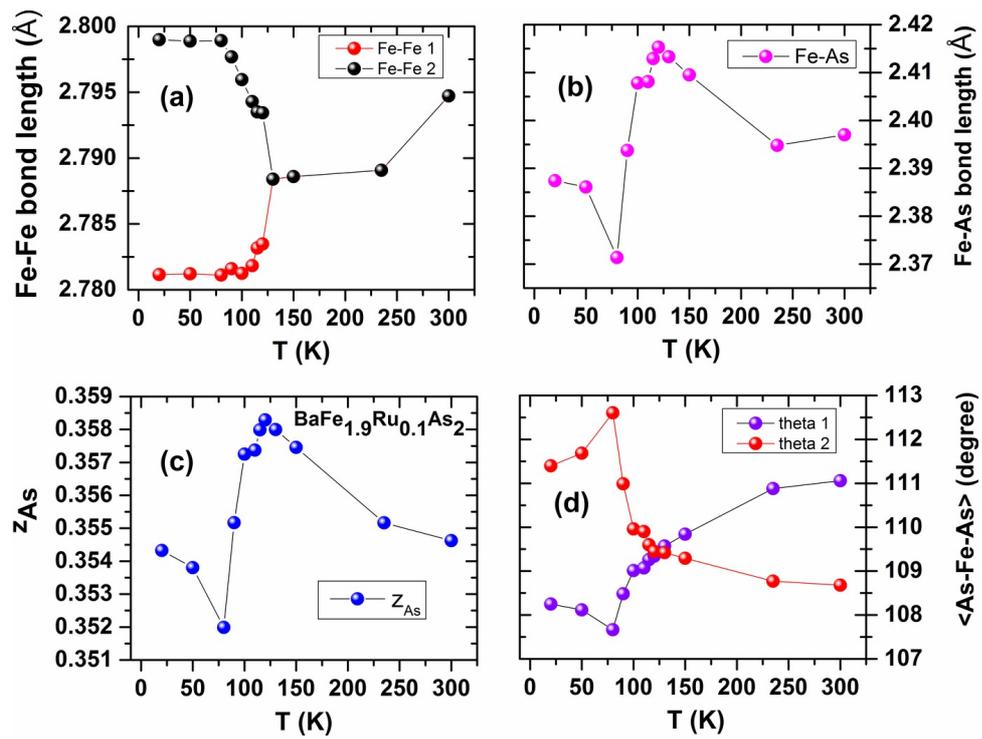

**Figure 6**

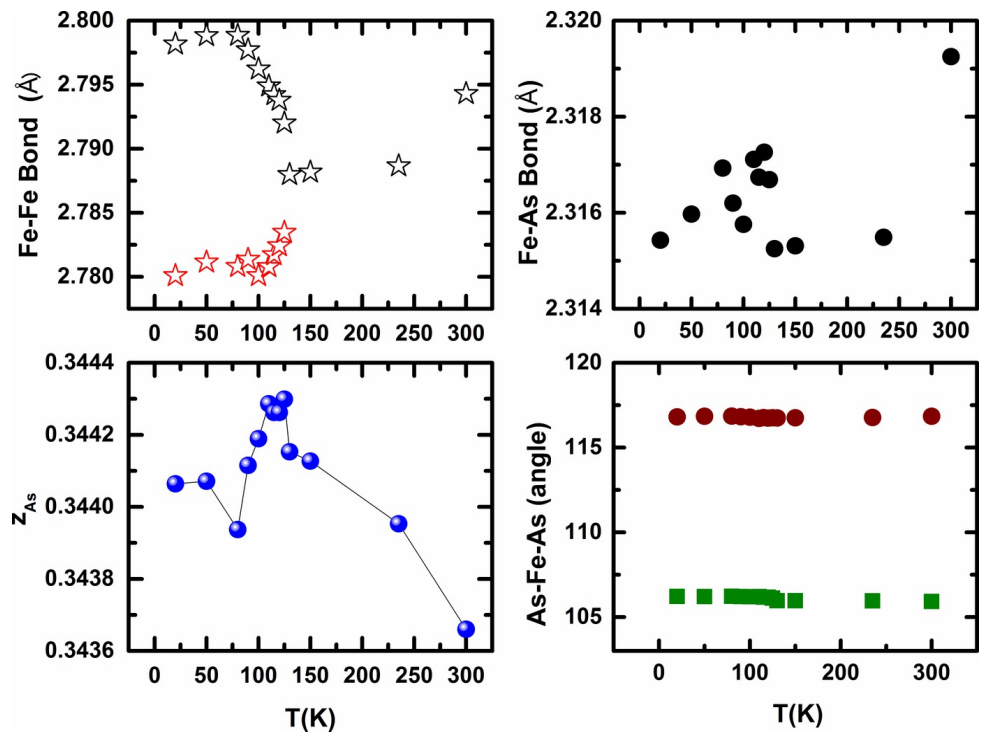

**Figure 7**

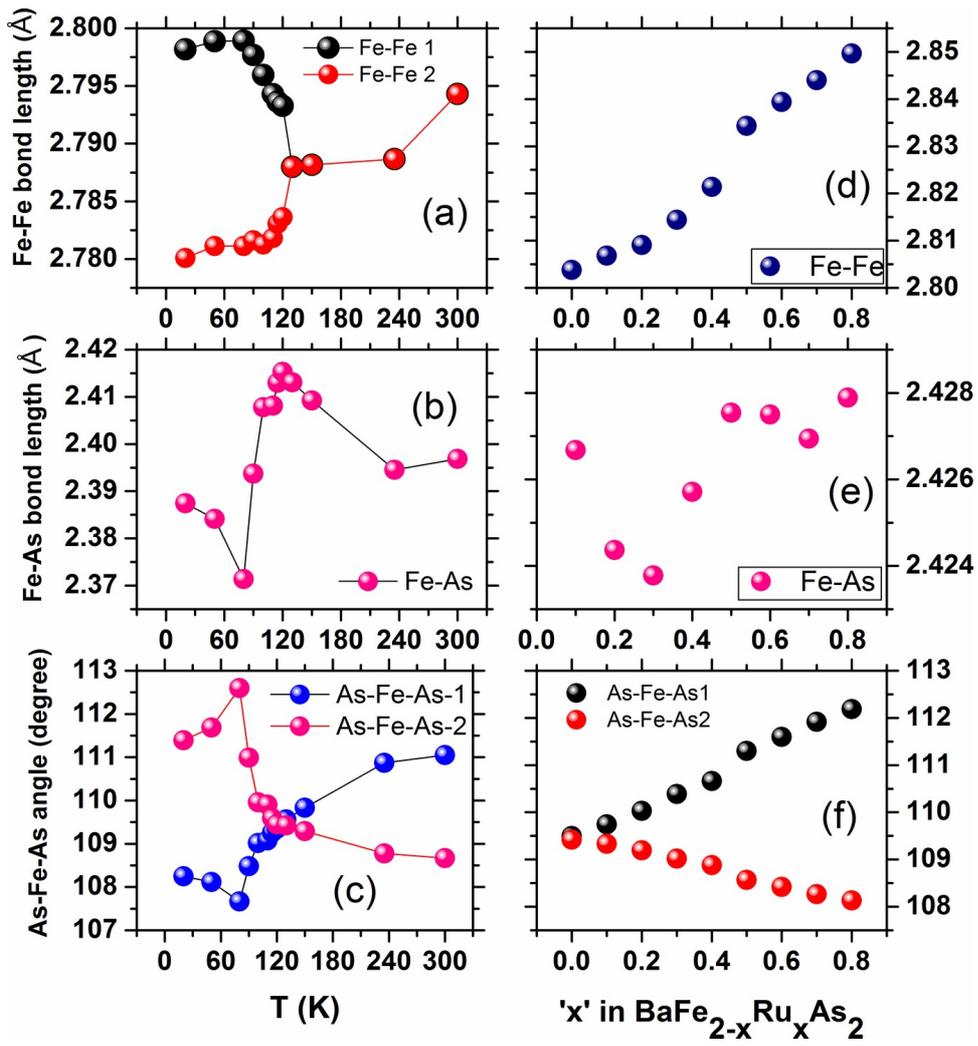

**Figure 8**